\documentclass[aps,prd,nofootinbib,twocolumn,superscriptaddress,longbibliography]{revtex4-1}
\usepackage{amssymb,amstext,amsmath,amsthm}
\usepackage{graphicx}
\usepackage{latexsym}
\usepackage{amsfonts}
\usepackage{verbatim}
\usepackage{color}  %needed to process latex in figures
\usepackage{amsmath,amsxtra,amsthm,amssymb,xr}
\usepackage[]{mathrsfs}
\newcommand{\bea}{\begin{eqnarray}}
\newcommand{\eea}{\end{eqnarray}}
\newcommand{\nn}{\nonumber}
\newcommand{\h}{\hspace{1mm}}

\newcommand{\C}{\mathbb{C}}

\newcommand{\CP}{\mathbb{CP}}
\newcommand{\SO}{{SO}}
\newcommand{\SL}{{SL}}
\newcommand{\U}{{U}}

\newcommand{\dd}{{\mathrm d}}  % exterior derivative (\d already defined)
\newcommand{\bn}{{\vec{n}}} % bold font n vector
\newcommand{\bz}{{\bf z}}

\def\la{\langle}
\def\ra{\rangle}
\newcommand{\ip}[1]{\la {#1}\ra}

\newcommand{\zn}[1]{\| Z_{#1}\|}
\newcommand{\jx}{J\xi_{ab}}
\newcommand{\cohe}{\ket{j,\Upsilon(\vec{n})}}

\def\nb{\mathbf{n}}

\newcommand{\bra}[1]{\la{#1}|}
\newcommand{\ket}[1]{|{#1}\ra}
\def\tot{S_\mathrm{tot}}
\def\zb{\bar{z}}
\def\h{H_{(ai)(aj)}}
\def\d{\mathrm{d}}
\newcommand{\rep}{{\cal H}}  % Lorentz group representation
\newcommand{\SLtwoC}{\SL(2,\mathbb{C})}
\newcommand{\slc}{\SL(2,\mathbb{C})}
\newcommand{\half}{\frac{1}{2}}
\newcommand{\z}{Z}  %some notation which might vary

\newcommand{\mc}[1]{\mathcal{#1}}

\newcommand{\mbb}[1]{\mathbb{#1}}

\newcommand{\mdots}{,.\,.\,,}
\newcommand{\Ref}[1]{(\ref{#1})}
\newcommand{\no}{\nonumber\\}
\newcommand{\eqa}{\begin{eqnarray}}
\newcommand{\neqa}{\end{eqnarray}}
\newcommand{\be}{\begin{equation}}
\newcommand{\ee}{\end{equation}}
\def\be{\begin{eqnarray}}
\def\ee{\end{eqnarray}}

\newcommand{\cp}{\mathcal P}

\renewcommand{\d}{\delta}

\newcommand{\SU}{{SU}}

\def\ra{\rangle}
\def\la{\langle}

\def\d{\delta}
\newcommand{\Hil}{\mathcal{H}}
\begin{document}
%\title{\Large  \textbf {Graviton propagator from Lorentzian spin foam}}
\title{Lorentzian spinfoam propagator}
\author{Eugenio Bianchi}
\affiliation{Centre de Physique Th\'eorique de Luminy, Case 907, F-13288 Marseille, France}
\thanks{Unit\'e mixte de recherche du CNRS et des Universit\'es de Provence, de la M\'editerran\'ee et du Sud; affili\'e \`a la FRUMAN.}
\affiliation{Perimeter Institute, 31 Caroline St N, Waterloo ON, Canada N2L 2Y}
\author{You Ding}
\affiliation{Centre de Physique Th\'eorique de Luminy, Case 907, F-13288 Marseille, France}
\thanks{Unit\'e mixte de recherche du CNRS et des Universit\'es de Provence, de la M\'editerran\'ee et du Sud; affili\'e \`a la FRUMAN.}
\affiliation{Department of Physics, Beijing Jiaotong University, Beijing 100044 , China}
\date{\small\today}

\begin{abstract}
The two-point correlation function is calculated in the Lorentzian EPRL spinfoam model, and shown to match with the one in Regge calculus in a proper limit: large boundary spins $j\to \infty$, and small Barbero-Immirzi parameter $\gamma\to 0$, keeping the size of the quantum geometry $A\sim \gamma j$ finite and fixed. Compared to the Euclidean case, the definition of a Lorentzian boundary state involves a new feature: the notion of past- and future-pointing intertwiners. The semiclassical correlation function is obtained for a time-oriented semiclassical boundary state.
\end{abstract}
\maketitle

\section{Introduction}
Spinfoam amplitudes provide a covariant definition of the dynamics of Loop Quantum Gravity \cite{carlo,tman,Han:2005km,Rovelli:2011eq}. A basic test the amplitude has to pass is that, its semiclassical behavior matches with classical gravity. The vertex amplitude has been shown \cite{Conrady:2008mk,Barrett:2009gg,Barrett:2009mw,Barrett:2009cj,Magliaro:2011zz,Magliaro:2011dz,Han:2011rf,Han:2011re} to correctly determine the Regge action for discrete gravity.  Given this result, the next test for the theory regards the behavior of small quantum fluctuations around a classical solution. In the Euclidean theory, the two-point correlation function for the Penrose metric operator (also called \emph{the spinfoam propagator}) restricted to a single vertex is calculated in \cite{Bianchi:2009ri} and the three-point function in \cite{rovelli:2011kf}. In this paper we will consider the physically relevant Lorentzian case.

In this paper
we compute the Lorentzian two-point correlation functions for the Penrose metric operator. The setting is the one introduced in \cite{Rovelli:2005yj} and developed in \cite{Bianchi:2006uf,Livine:2006it,Bianchi:2007vf,Christensen:2007rv,Alesci:2007tx,Alesci:2007tg,Alesci:2008ff,Speziale:2008uw,Bianchi:2009ri,Neiman:2011gf}. In particular, we restrict to a single spinfoam vertex and compute correlations on a semiclassical state peaked on the spacelike boundary geometry of a Lorentzian $4$-simplex. For a discussion of the role of this truncation see \cite{Rovelli:2011mf}.

Here we briefly review the procedure for completeness. We consider a 4-dimensional manifold $\mc{R}$ homeomorphic to a $4$-ball, with its boundary given by  a $3$-dimensional manifold $\Sigma=\partial \mc{R}$ homeomorphic to the $3$-sphere $S^3$. We associate to $\Sigma$ a boundary Hilbert space of states: the LQG Hilbert space $\mc{H}_\Sigma$ spanned by (abstract) spin networks. We call $|\Psi\rangle$ a generic state in $\mc{H}_\Sigma$. A spin foam model for the region $\mc{R}$ provides a map from the boundary Hilbert space to $\mbb{C}$.  We call this map $\langle W|$. It provides a sum over the bulk geometries with a weight that defines our model for quantum gravity. The dynamical expectation value of an operator $\mc{O}$ on the state $|\Psi\rangle$ is defined via the following expression
\begin{equation}
\langle \mc{O}\rangle=\frac{\langle W| \mc{O}|\Psi\rangle}{\langle W|\Psi\rangle}\;.
%\label{eq:}
\end{equation}
{This expression corresponds to the standard definition in (perturbative) quantum field theory where the \emph{vacuum} expectation value of a product of local observables is defined as
\begin{align}
\langle O(x_1)\cdots& O(x_n)\rangle_0=\frac{\displaystyle \int D[\varphi] O(x_1)\cdots O(x_n) e^{i S[\varphi]}}{\displaystyle \int D[\varphi]  e^{i S[\varphi]}}\no&\equiv \frac{\displaystyle \int D[\phi] W[\phi] O(x_1)\cdots O(x_n) \Psi_0[\phi]}{\displaystyle \int D[\phi] W[\phi] \Psi_0[\phi]}\;.
%\label{eq:}
\end{align}
The vacuum state $\Psi_o[\phi]$ codes the boundary conditions at infinity.}
The operator $\mc{O}$ can be a geometric operator as the area, the volume or the length \cite{Rovelli:1994ge,Ashtekar:1996eg,Ashtekar:1997fb,Ashtekar:1998ak,Major:1999mc,Thiemann:1996at,Bianchi:2008es,Ma:2010fy}. The geometric operator we are here interested in is the (density-two inverse-) metric operator $q^{ab}(x)=\delta^{ij}E^a_i(x) E^b_j(x)$. We focus on the \emph{connected} two-point correlation function $G^{abcd}(x,y)$ on a semiclassical boundary state $|\Psi_0\rangle$. It is defined as
\begin{equation}
G^{abcd}(x,y)=\langle q^{ab}(x)\; q^{cd}(y)\rangle - \langle q^{ab}(x)\rangle\, \langle q^{cd}(y)\rangle\;.
\label{eq:G}
\end{equation}

We generalize the semiclassical states used in \cite{Bianchi:2009ri}, which is obtained via a superposition over spins of spin-network states having nodes labeled by Livine-Speziale coherent intertwiners \cite{Livine:2007vk,Conrady:2009px,Freidel:2009nu}, with coefficients proposed by Rovelli in \cite{Rovelli:2005yj}. The origins of this semiclassical state is studied in \cite{Bianchi:2009ky}. Here we generalize this semiclassical state to be Lorentzian, in the sense that it is related to Lorentzian geometry.

We assume that the Barbero-Immirzi parameter $\gamma$ is positive \footnote{
In \cite{Neiman:2011gf}, it is pointed out that both of the vertex amplitude and the correlation function are invariant under $\gamma\rightarrow -\gamma$.
} and restrict ourselves to the first order of vertex expansion, i.e., we consider the spinfoam with a single vertex.
Our main result is the following. We consider the limit, introduced in \cite{Bianchi:2009ri} and discussed in  \cite{Magliaro:2011dz,Magliaro:2011zz}, where the Barbero-Immirzi parameter is taken to zero $\gamma\to 0$, and the spin of the boundary state is taken to infinity $j\to \infty$, keeping the size of the quantum geometry $A\sim \gamma j$ finite and fixed. This limit corresponds to neglecting Planck scale discreteness  and twisting effects, at large finite distances. In this limit, the two-point function we obtain exactly matches the one obtained from Lorentzian Regge calculus \cite{Regge:1961px}. We therefore extend to Lorentzian signature the results of \cite{Bianchi:2009ri}.

\section{Lorentzian geometry and Lorentzian semiclassical boundary states}\label{sec:reversal}
Semiclassical boundary states are a key ingredient in the definition of boundary amplitudes. Here we describe in detail the construction of a boundary state peaked on the intrinsic and the extrinsic geometry of the boundary of a Lorentzian $4$-simplex. The construction follows the Euclidean case \cite{Bianchi:2009ri}, by considering Lorentzian geometry of a $4$-simplex with space-like tetrahedra. As in the Euclidean case \cite{Bianchi:2009ri}, it uses the Livine-Speziale coherent intertwiners  \cite{Livine:2007vk} (see also \cite{Conrady:2009px}) together with a superposition over spins as done in \cite{Rovelli:2005yj,Bianchi:2006uf}, and thus it can be considered as a Lorentzian version of an improvement of the boundary state used in \cite{Alesci:2007tx,Alesci:2007tg,Alesci:2008ff} where Rovelli-Speziale gaussian states \cite{Rovelli:2006fw} for intertwiners were used.

We consider a simplicial decomposition $\Delta_5$ of $S^3$. The decomposition $\Delta_5$ is homeomorphic to the boundary of a Lorentzian $4$-simplex: it consists of five tetrahedra $a$ which meet at ten triangles ${(ab)}$ ($a,b=1...5$ and $a<b$). In particular,we focus on the Lorentzian geometry, where all the five tetrahedra $a$ are space-like, i.e., all the normals to the tetrahedra are time-like. Suppose the time-like normals to the tetrahedra are outward-pointing, for convenience. Then for each triangle $(ab)$, there is a corresponding ``wedge'' composed by the two tetrahedra $a$ and $b$ which meet at the triangle $(ab)$. Since all the tetrahedra are space-like, the tetrahedron comes in two types:  the outward normals are either future-pointing or past-pointing. The wedges are then classified into two types: it is called in \cite{Barrett:1993db,Barrett:2009mw} \emph{thick wedge} if the incident tetrahedra are of same pointing type, which means both future-pointing or both past-pointing, otherwise called \emph{thin wedge}. We introduce a quantity $\Pi_{ab}$ to denote the Lorentzian geometry of the triangle $(ab)$ in the following way:
\begin{eqnarray}
\Pi_{ab}=\left\{%
\begin{array}{ll}
    0, & \hbox{thick wedge} \\
    \pi, & \hbox{thin wedge} \\
    \end{array}%
\right.\label{eq:Pi}
\end{eqnarray}

Now we consider the complete graph $\Gamma_5$ dual to the decomposition $\Delta_5$:
\begin{equation}
\Gamma_5=\;\;\parbox[c]{130pt}{\includegraphics[width=5cm]{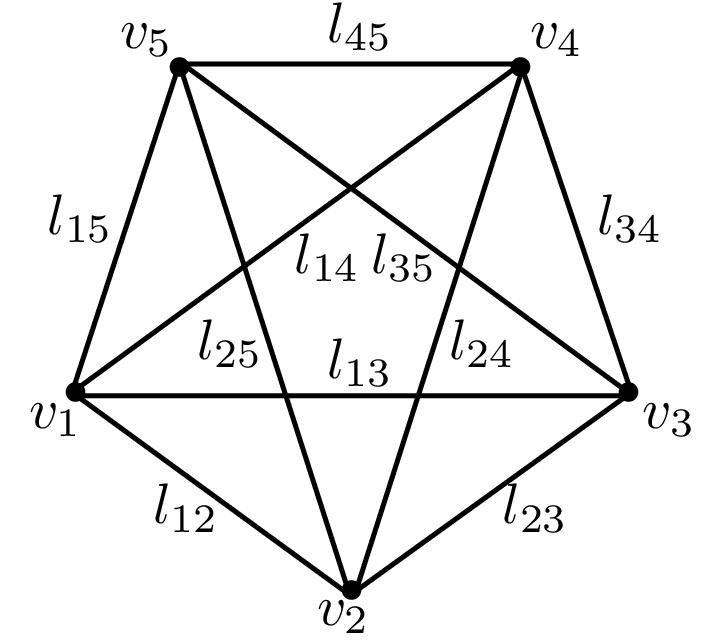}}\;,
\label{eq:G5}
\end{equation}
with five nodes $v_a$ dual to tetrahedra of the 4-simplex and ten links $l_{ab},\;(a<b)$ dual to the corresponding meeting triangles.
Consider the group $SU(2)$ and associate
an irreducible representation
$j_l$ (a spin) to each link $l$ of the graph $\Gamma$, and
 an $SU(2)$ intertwiner $i_n$
to each
node $n$ of the graph. The \emph{spin network states} $\ket{\Gamma_5,j_{ab},i_a}$ supported on this graph span the truncated $SU(2)$ Hilbert space $\Hil_{\Gamma_5}$ of LQG.

Now let us introduce a (overcomplete) basis of $\Hil_{\Gamma_5}$,which is the \emph{coherent spin network states} with nodes labeled by Livine-Speziale coherent intertwiners. We start from associating
  an \emph{$SU(2)$ coherent state} $\ket{j,\vec{n}(\xi)}$ \cite{Livine:2007vk,coherent} in the spin-$j$ representation to each link of the graph $\Gamma_5$, labeled by the spin-$j$ and a unit vector $\vec{n}(\xi)$ defining a direction on the sphere $S^2$, associated to a normalized spinor $\xi\in\C^2$. In fact, one can obtain these coherent states from the maximal weight vectors $\ket{j,j}$, when $m=j$ on the basis states $\ket{j,m}$, which minimize the ($\SU(2)$ invariant) uncertainty $\Delta\equiv\,|\la\vec{J}^2\ra-\la\vec{J}\ra^2|$ in the direction of $J_z$.
Starting from the highest weight, an infinite set of coherent states
on the sphere $\SU(2)/\U(1)\sim {S}^2$ are constructed through the group action,
$
\ket{j,\vec{n}} =n\ket{j,j},
$
where $\vec{n}$ is a unit vector defining a direction on the
sphere ${S}^2$ and $n$ an $SU(2)$ group element
rotating the direction $\hat{z}\equiv (0,0,1)$ into the direction
$\vec{n}$.
Just as $\ket{j,j}$ has direction $z$ with minimal uncertainty, $\ket{j, \vec{n}}$ has direction $\vec{n}$
with minimal uncertainty. One can go further to use a normalized spinor $\xi\in \C^2$ to label an $SU(2)$ group element
\be
n(\xi)=\left(\begin{array}{cc} \xi_0 & -\bar{\xi}_1 \\
                           \xi_1 & \bar{\xi}_0  \end{array}\right) \nonumber
\ee
and the corresponding vector $\vec{n}(\xi)$.

These coherent states  are peaked on the geometry of a classical triangle: $\vec{n}$ are associated to unit-normals to triangles of a tetrahedron, and $j$ areas of the triangles. We assume all the normals outward to the tetrahedron, which satisfy
\begin{align}
j_1\vec{n}_1+j_2\vec{n}_2+j_3\vec{n}_3+j_4\vec{n}_4=0 \label{outward},
\end{align}
thus we associate to each triangle $(ab)$ normals $-\vec{n}_{ab}$ and $\vec{n}_{ba}$ when  $a$ is target of the triangle and respectively $b$ is the source. The antipodal vector $-\vec{n}(\xi)$ can be associated to $J\xi$, i.e. $-\vec{n}(\xi)=\vec{n}(J\xi)$, with
\be
J\begin{pmatrix}\xi_0\\\xi_1\end{pmatrix}=\begin{pmatrix}-\bar{\xi}_1\\\ \ \bar{\xi}_0\end{pmatrix}.\label{Jxi}
\ee

A Euclidean coherent intertwiner between the representations $j_1\mdots j_4$ is defined as
\begin{equation}
\Phi(\vec{n}_1\mdots \vec{n}_4)=\int_{SU(2)}dh \; \prod_{a=1}^{4}\,\langle j_a,m_a| h |j_a,\vec{n}_a\rangle
\label{eq:eu intertwiner}
\end{equation}
up to a normalization, with the $SU(2)$ coherent states $\ket{j,\vec{n}}$ satisfying the closure constraint (\ref{outward}). Here we use $h$ to denote the $SU(2)$ group element as well as the corresponding representation. Let us now introduce the notion of \emph{Lorentzian Livine-Speziale intertwiner}, which generalizes the Euclidean Livine-Speziale intertwiner \cite{Livine:2007vk} to the one related to the Lorentzian geometry.
Since one has the closure condition for the 4-simplex and all the tetrahedra are space-like, with outward normals,  there is at least one (and at most four) tetrahedron with past-pointing outward normal. The Lorentzian coherent intertwiner $\Upsilon(\vec{n}_1\mdots \vec{n}_4)$ can be obtained from the Euclidean one (\ref{eq:eu intertwiner}) by time reversing the normals of past-pointing tetrahedra:
\begin{itemize}
  \item For future-pointing tetrahedra, \be\Upsilon(\vec{n}_1\mdots \vec{n}_4)=\Phi(\vec{n}_1\mdots \vec{n}_4);\label{upsilon:future}\ee
  \item For past-pointing tetrahedra, \be\Upsilon(\vec{n}_1\mdots \vec{n}_4)=T\Phi(-\vec{n}_1\mdots -\vec{n}_4).\label{upsilon:past}\ee
\end{itemize}
The effect of time reversal $T$ on the coherent states are given by
\be
T\ket{j,\vec{n}}=(-1)^j\ket{j,-\vec{n}},
\ee
and thus on the (past-pointing) coherent intertwiner by
\begin{align}
T\Phi_a(\vec{n}_1\mdots\vec{n}_4)=(-1)^{\sum_{b> a}j_{ab}}\Phi_a(-\vec{n}_1\mdots-\vec{n}_4).\label{time reversal}
\end{align}
Here we assume the normal to the 5th tetrahedron are future-pointing and thus $a=1,...4.$
The minus signs in the r.h.s. of equation (\ref{upsilon:past}) keeps normals outward-pointing.
%In fact, the action of time reversal $T$ changes the signs of the normals $\vec{n}$ (shown in equation \Ref{time reversal}), which breaks the gluing %condition
%% 12/08/2011 discuss
Combining the two components \Ref{upsilon:future} and \Ref{upsilon:past}, one obtains
\begin{align}
\Upsilon_a(\vec{n}_1\mdots\vec{n}_4)=\exp{(-i\sum_{b>a}\Pi_{ab}j_{ab})}\Phi_a(\vec{n}_1\mdots\vec{n}_4),\label{intertwiner relation}
\end{align}
with $\Pi_{ab}$ related to Lorentzian geometry in equation \Ref{eq:Pi}. From the intertwiner relation (\ref{intertwiner relation}), one can call $\exp{(-i\sum_{a<b}\Pi_{ab}j_{ab})}$ \emph{Lorentzian-geometry phase}, which maps the Euclidean coherent intertwiners to the Lorentzian ones.

This Lorentzian coherent intertwiner can be used to label nodes of \emph{Lorentzian coherent spin network state}.
Calling $v_i^{m_1\cdots m_4}$ the standard recoupling basis for intertwiners, we can define the coefficients
\begin{equation}
\Upsilon_i(\vec{n}_1\mdots \vec{n}_4)=v_i^{m_1\cdots m_4} \Upsilon_{m_1\cdots m_4}(\vec{n}_1\mdots \vec{n}_4)\;.
%\label{eq:}
\end{equation}
We define a \emph{Lorentzian coherent spin network} $\ket{\Gamma_5,j_{ab},\Upsilon_a}$ as the state labeled by ten spins $j_{ab}$ and $4\times 5$ normals $\vec{n}_{ab}$ and given by the superposition
\begin{equation}
|\Gamma_5,j_{ab},\Upsilon_a(\vec{n})\rangle= \sum_{i_1\cdots i_5}\,\Big(\prod_{a=1}^{5} \Upsilon_{i_a}(\vec{n}_{ab})\Big) |\Gamma_5,j_{ab},i_a\rangle\;.
%\label{eq:}
\end{equation}
If one uses $\ket{\Gamma_5,j_{ab},\Phi_a(\vec{n})}$ to denote the Euclidean coherent spin network state with nodes labeled by Euclidean coherent intertwiner $\Phi_a(\vec{n})$, as in \cite{Bianchi:2009ri}, the Lorentzian coherent spin network state $\ket{\Gamma_5,j_{ab},\Upsilon_a(\vec{n})}$ can be expressed as $\ket{\Gamma_5,j_{ab},\Phi_a(\vec{n})}$ times a Lorentzian-geometry phase:
\begin{align}
\ket{\Gamma_5,j_{ab},\Upsilon_a(\vec{n})}=\exp{(-i\sum_{a<b}\Pi_{ab}j_{ab})}\ket{\Gamma_5,j_{ab},\Phi_a(\vec{n})}.\label{spin network relation}
\end{align}
The two phases from the intertwiner relation \Ref{intertwiner relation} and from the spin network relation \Ref{spin network relation} are different, in the sense that $a$ is fixed in the former but free in the latter. In fact, the latter phase is product of the five phase in the former form. However, here, we call both of them \emph{Lorentzian-geometry phase}, if there is no confusion rises.

A Lorentzian semiclassical state peaked both on intrinsic and extrinsic geometry can be given by a superposition of Lorentzian coherent spin network states:
\begin{equation}
|\Psi_o\rangle=\sum_{j_{ab}} \psi_{j_o,\phi_o}(j) |j,\Upsilon(\bn)\rangle\;,
\label{semiclassical state}
\end{equation}
with coefficients $\psi_{j_o,\phi_o}(j)$ given by a gaussian times a phase,
\begin{align}
&\psi_{j_o,\phi_o}(j)=\exp\Big({-i\sum_{ab}\gamma \phi_o^{ab}\,(j_{ab}-(j_o)_{ab})}\Big)\times\no
&\times\exp\Big(-\sum_{ab,cd} \gamma\alpha^{(ab)(cd)}\, \frac{j_{ab}-{(j_o)}_{ab}}{\sqrt{(j_o)_{ab}}}\,\frac{j_{cd}-(j_o)_{cd}}{\sqrt{(j_o)_{cd}}}\Big)\;.
\label{smallpsi}
\end{align}
where $\phi_o$ labels the simplicial extrinsic curvature, which is an angle associated to the triangle shared by the tetrahedra; the $10\times 10$ matrix $\alpha^{(ab)(cd)}$ is assumed to be complex with positive definite real part. These coefficients are originally proposed in \cite{Rovelli:2005yj} and used to calculate the Euclidean two-point \cite{Bianchi:2009ri}  and three-point \cite{rovelli:2011kf} correlation functions. Here we will use the Lorentzian semiclassical state \Ref{semiclassical state} to calculate the Lorentzian two-point correlation function.

\section{The EPRL amplitude in Lorentzian theory}\label{eprl}
In this section, we give a brief introduction to the $\slc$ EPRL amplitude of a coherent spin network.
Throughout this paper, $\SL(2,\C)$ refers to the 6-dimensional real Lie group of $2\times2$ complex matrices with unit determinant, and is called simply the Lorentz group. It covers the group of proper orthochronous Lorentz transformations, $\SO^+(3,1)$, which is the component of the group $O(3,1)$ connected to the identity.

The principal series of irreducible unitary representations of the Lorentz group $\SLtwoC$
are labeled by two parameters $(k, p)$, with $k$ an integer and $p$ a real number \cite{gms}. Given a carrier space $\rep_{(k,p)}$, the canonical basis is given by the basis diagonalizing simultaneously the Casimir operators, which is denoted as $\ket{(k,p);j,m}$.

The $\SLtwoC$ EPRL amplitude of a single 4-simplex for a Lorentzian coherent spin network state $\ket{\Gamma_5,j_{ab},\Upsilon_a(\vec{n})}$ reads
\begin{align}
&\bra{W}\Gamma_5,j_{ab},\Upsilon_a(\vec{n})\ra\no
&=e^{-i\sum_{ab}\Pi_{ab}j_{ab}}\int\limits_{SL(2,\mathbb{C})^{5}}\prod_{a}\mathrm{d}g_a\prod_{(ab)}P_{ab}{(g)},\label{amplitude bracket}
\end{align}
with
\begin{align}
P_{ab}(g)=\langle j_{ab},-\vec{n}_{ab}(\xi)|\,Y^\dag g_a^{-1} g_b Y |j_{ab},\vec{n}_{ba}(\xi)\rangle.\label{Pab bracket}
\end{align}
Notation is as follows.
The indices $a,b=1,...,5$ label the tetrahedra on the boundary of the 4-simplex and $(ab)$ labels the triangles between the corresponding tetrahedra; the integral is over one group element of $\slc$ per each tetrahedron. We restrict ourselves to the spacelike tetrahedra. We use $g$ to denote the group elements, as well as the corresponding representations. The EPRL embedding map $Y$ embeds the spin-$j$ irreducible representation $\Hil_j$ of $SU(2)$ to the irreducible unitary representation $\Hil_{(k,p)}$ of $\slc$,  given by
\be
Y\ket{j,m}=\ket{(j,\gamma j);j,m}.
\ee
Note that we associate to each triangle $(ab)$ normals $-\vec{n}_{ab}$ and $\vec{n}_{ba}$ when  $a$ is target of the triangle and respectively $b$ is the source. That is why we have a minus sign in the bra coherent state in the definition of $P_{ab}$ in equation (\ref{Pab bracket}).
 And the Hermitian inner product is defined as
$
\ip{z,w}=\bar{z}_ow_o+\bar{z}_1w_1
$.

To see this amplitude (\ref{amplitude bracket}) explicitly, one can turn to a representation of the Lorentz group $\slc$ on the space $\rep_{(k,p)}$ of homogeneous functions of the \emph{complex affine plane} $\C^2-\{0,0\}$,
\begin{align}
f(a\bz)^{(k,p)}&=a^{-1+ip+k}\bar{a}^{-1+ip-k}f(\bz)^{(k,p)}, \no
&\forall a\in \C-\{0\},\label{homofunc}
\end{align}
with the group transformation
\be
g:\ f(\bz)\mapsto f(g^T\bz).
\ee
The
canonical basis is denoted as $f^j_m(\bz)^{(k,p)}$.
And the inner product is given by
\be
(f,g)=\int\dd\bz\,\bar{f}g
\ee
with $\dd\bz\equiv\frac i2(z_0\dd z_1-z_1\dd z_0)\wedge (\bar z_0\dd \bar z_1-\bar z_1\dd \bar z_0)$. This integral is invariant under the scaling
$\bz\rightarrow a\bz$, according to the homogeneity (\ref{homofunc}). To modulo this equivalence relation, one can choose
$
\varphi(z)=f(z,1)
$
associated with each $f(z_o,z_1)\in\rep_{(k,p)}$; the functions $\varphi(z)$ forms a realization of $\rep_{(k,p)}$, which we can still call $\rep_{(k,p)}$. Functions $\varphi(z)$ can be considered as the homogeneous functions on the {\it complex projective line} $\cp$, the subspace of the complex affine plane,
modulo the equivalence relation $a\bz=\bz$. For calculating simplicity, we will still keep the formulae of $f(\bz)$ on the complex affine plane in the following and reduce to $\varphi(z)\in\cp$ when necessary.

In this representation of homogeneous functions, $\slc$ coherent state $\ket{(k,p);k,\vec{n}(\xi)}$  with lowest spin $k$ can be written  as \cite{Barrett:2009mw}
\begin{align}
f^k_\xi(z)^{(k,p)}= \sqrt{\frac{d_k}{\pi}}\, \la z, z\ra^{ip-1-k}\; \la\bar{z} ,\xi \ra^{2k}.\label{fkxi}
\end{align}
And hence equation (\ref{Pab bracket}) can be rewritten \cite{roberto_thesis} as
\begin{widetext}
\begin{align}
P_{ab}=&\langle j_{ab},-\vec{n}_{ab}(\xi)|\,Y^\dag \,g_a^{-1} g_b\, Y |j_{ab},\vec{n}_{ba}(\xi)\rangle\nonumber\\
=&\langle (j_{ab},\gamma j_{ab}); j_{ab},-\vec{n}_{ab}(\xi)|\,g_a^{-1}g_b\,  |(j_{ab},\gamma j_{ab});j_{ab},\vec{n}_{ba}(\xi)\rangle\nonumber\\
=&\int_{\CP^1}\dd\bz\;\overline{g_a\,f^{j_{ab}}_{J\xi_{ab}}(z)^{(j_{ab},\gamma j_{ab})}}\;g_b\,f^{j_{ab}}_{\xi_{ba}}(z)^{(j_{ab},\gamma j_{ab})}  \nonumber\\
=&\int_{\CP^1}\dd\bz\;\overline{f^{j_{ab}}_{J\xi_{ab}}(g^T_az)^{(j_{ab},\gamma j_{ab})}}\;f^{j_{ab}}_{\xi_{ba}}(g^T_bz)^{(j_{ab},\gamma j_{ab})}  \nonumber\\
=&\frac{d_{j_{ab}}}{\pi}\int_{\CP^1}\dd\bz\;\la g_a^{\dagger}\bar{z},g_a^{\dagger}\bar{z}\ra^{-1-(1+i\gamma)j_{ab}}\la J\xi_{ab},g_a^{\dagger}\bar{z}\ra^{2j_{ab}}
\la g_b^{\dagger}\bar{z},g_b^{\dagger} \bar{z}\ra^{-1-(1-i\gamma)j_{ab}}\la g_b^{\dagger}\bar{z}, \xi_{ba},\ra^{2j_{ab}}  \nonumber\\
=&-\frac{d_{j_{ab}}}{\pi}\int_{\CP^1}\dd\bz\;\la g_a^{\dagger}{z},g_a^{\dagger}{z}\ra^{-1-(1+i\gamma)j_{ab}}\la J\xi_{ab},g_a^{\dagger}{z}\ra^{2j_{ab}}
\la g_b^{\dagger}{z},g_b^{\dagger} {z}\ra^{-1-(1-i\gamma)j_{ab}}\la g_b^{\dagger}z, \xi_{ba},\ra^{2j_{ab}}  \nonumber\\
=&\frac{d_{j_{ab}}}{\pi}\int_{\CP^1}\dd\tilde{\bz}_{ab}\;\left(\frac{\la Z_{ba},Z_{ba}\ra}{\la Z_{ab},Z_{ab}\ra}\right)^{i\gamma j_{ab}}\left(\frac{\la J\xi_{ab},Z_{ab}\ra^2\la Z_{ba},\xi_{ba}\ra^2}{\la Z_{ab},Z_{ab}\ra \la Z_{ba},Z_{ba}\ra}\right)^{j_{ab}},\label{Pab_calculate}
\end{align}
\end{widetext}
where $\dd \tilde{\bz}_{ab}\equiv-(\la Z_{ab},Z_{ab}\ra \la Z_{ba},Z_{ba}\ra)^{-1}\dd \bz$, $Z_{ab} \equiv g_a^{\dag}z$ and $Z_{ba} \equiv g_b^{\dag}z$; $\xi_{ba}$ and $J\xi_{ab}$ are spinors associated respectively with $\vec{n}_{ba}(\xi)$ and $-\vec{n}_{ab}(\xi)$, as introduced in equation (\ref{Jxi}); note that $g$ is used to denote the group elements, as well as the corresponding unitary representations; the property of unitary representation is considered in the 3rd step ,equation (\ref{fkxi}) used in the 5th step and in the 6th step, the integral variable $z$ is changed into its complex conjugate $\bar{z}$, where the minus sign comes from the integral measure.
If we denote the integrand in equation \Ref{Pab_calculate} as $K_{ab}$:
\begin{align}
K_{ab}&(g,\bz)=\left(\frac{\la Z_{ba},Z_{ba}\ra}{\la Z_{ab},Z_{ab}\ra}\right)^{i\gamma j_{ab}}\times\no
&\times\left(\frac{\la J\xi_{ab},Z_{ab}\ra^2\la Z_{ba},\xi_{ba}\ra^2}{\la Z_{ab},Z_{ab}\ra \la Z_{ba},Z_{ba}\ra}\right)^{j_{ab}},\label{Kab}
\end{align}
the integral \eqref{Pab_calculate} is simply expressed as
\be
P_{ab}=\int\frac{d_{j_{ab}}}{\pi}\dd\tilde{\bz}_{ab}\;K_{ab}.
\ee
Thus
the EPRL amplitude (\ref{amplitude bracket}) can be written as
\begin{align}
&\ip{W|\Gamma_5,j_{ab},\Upsilon_a(\bn)}\no
&=\int\prod_{a=1}^5\dd g_a\, \int \left(\prod_{a<b}\;\frac{d_{j_{ab}}}{\pi}\dd \tilde{\bz}_{ab}\right)\,e^{S},\label{amplitude integral}
\end{align}
where the ``action" $S$ is given by
\begin{align}
S(g,\bz)=\sum_{a<b}&\left(j_{ab} \log \frac{\la J\xi_{ab},Z_{ab}\ra^2\la Z_{ba},\xi_{ba}\ra^2}{\la Z_{ab},Z_{ab}\ra
 \la
 Z_{ba},Z_{ba}\ra} \right.\no
&\left. + i  \gamma j_{ab} \log \frac{\la Z_{ba},Z_{ba}\ra}{\la Z_{ab},Z_{ab}\ra}-i\Pi_{ab}j_{ab}\right).\label{action}
\end{align}
Note that the ``action" $S$ here has one more term $-i\sum \Pi_{ab} j_{ab}$ than the one in \cite{Barrett:2009mw}, since we consider the \emph{Lorentzian} coherent spin network states.
 This expression, however, is ill defined, due to the fact that the integral may diverge.
This issue is addressed and answered in \cite{Baez:2001fh,Engle:2008ev}, where it is shown that
the source of the divergence is a redundant integral over $\slc$ in the vertex amplitude (\ref{amplitude integral}).
It is then immediate to regularize the vertex amplitude by removing one $\slc$ integration.
The resulting amplitude with an integral over $\prod_{a=1}^4\dd g_a$ is proven in  \cite{Baez:2001fh,Engle:2008ev} to be finite.

%-----------------------------------
%
%
%-----------------------------------
\section{Lorentzian two-point function and its integral formula}
Following \cite{Bianchi:2009ri}, the connected two-point correlation function $G^{abcd}_{nm}$ on a semiclassical boundary state $\ket{\Psi_o}$ is defined as
\begin{equation}
G_{nm}^{abcd}=\langle E_n^a\!\cdot\! E_n^b\; E_m^c\!\cdot\! E_m^d\rangle - \langle E_n^a\!\cdot\! E_n^b\rangle\, \langle E_m^c\!\cdot\! E_m^d\rangle\;,
\label{eq:Gnm}
\end{equation}
where $(E^a_n)_i$ is a flux operator through a surface $f_{an}$ dual to the triangle between the tetrahedra $a$ and $n$, parallel transported in the tetrahedron $n$.
Here the dynamical expectation value of an operator $\mathcal{O}$ on the semiclassical state $\ket{\Psi_o}$  is defined via
\begin{equation}
\langle \mathcal{O}\rangle=\frac{\langle W| \mathcal{O}|\Psi_o\rangle}{\langle W|\Psi_o\rangle}\;.\label{expectation}
\end{equation}
The semiclassical state $\ket{\Psi_o}$, which is {introduced in the end of section \ref{sec:reversal}}, can be simply expressed as a superposition:
\be
\ket{\Psi_o}=\sum_j\psi_j\ket{j,\Upsilon(\vec{n})},
\ee
with $\psi_j$ given by equation \Ref{smallpsi}.
Thus the two-point function (\ref{eq:Gnm}) can be also written as a superposition:
\begin{widetext}
\begin{align}
G_{nm}^{abcd}=&\frac{\sum_j\psi_j\bra{W}E_n^a\!\cdot\! E_n^b\; E_m^c\!\cdot\! E_m^d\cohe}{\sum_j\psi_j\la {W}\cohe}-\frac{\sum_j\psi_j\bra{W}E_n^a\!\cdot\! E_n^b\cohe}{\sum_j\psi_j\la {W}\cohe}\;
\frac{\sum_j\psi_j\bra{W} E_m^c\!\cdot\! E_m^d\cohe}{\sum_j\psi_j\la {W}\cohe}.\label{super:Gnm}
\end{align}
\end{widetext}

To see this explicitly, let us go first to derive integral expressions for $\bra{W}E^a_n\cdot E^b_n\ket{j,\Upsilon(\bn)}$ and $\bra{W}E^a_n\cdot E^b_n\ E^c_m\cdot E^d_m\ket{j,\Upsilon(\bn)}$.
As in \cite{Bianchi:2009ri}, one can introduce an ``insertion"
\begin{align}
Q_{ab}^i\equiv \langle j_{ab},-\vec{n}_{ab}(\xi)|\,Y^\dag g_a^{-1} g_b Y (E^a_b)^i |j_{ab},\vec{n}_{ba}(\xi)\rangle,\label{Qab bracket}
\end{align}
 and obtain the integral expression of $\bra{W}E^a_n\cdot E^b_n\ket{j,\Upsilon(\bn)}$ and $\bra{W}E^a_n\cdot E^b_n\ E^c_m\cdot E^d_m\ket{j,\Upsilon(\bn)}$ in terms of this insertion:
\begin{widetext}
\begin{align}
\langle W| E_n^{a}\!\cdot\! E_n^{b}|j,\Upsilon(\vec n)\rangle
&=\int \exp{(-i\sum_{ab}\Pi_{ab}j_{ab})}\prod_{a=1}^5 \dd g_a\, \delta_{ij} Q_{na}^i Q_{nb}^j {\prod_{cd}}' P^{cd}\;\no
\langle W| E_n^{a}\!\cdot\! E_n^{b}\;E_m^{c}\!\cdot\! E_m^{d}|j,\Upsilon(\vec n)\rangle
&=\int \exp{(-i\sum_{ab}\Pi_{ab}j_{ab})}\prod_{a=1}^5 \dd g_a\, \delta_{ij} Q_{na}^i Q_{nb}^j \delta_{kl} Q_{mc}^k Q_{md}^l {\prod_{cd}}' P^{cd}\;,
\label{eq:QQP}
\end{align}
\iffalse
\begin{align}
&\langle W| E_n^{a}\!\cdot\! E_n^{b}|j,\Upsilon(\vec n)\rangle\no
&=\int \prod_{a=1}^5 \dd g_a\, \delta_{ij} Q_{na}^i Q_{nb}^j {\prod_{cd}}' P^{cd}\;\no
&\langle W| E_n^{a}\!\cdot\! E_n^{b}\;E_m^{c}\!\cdot\! E_m^{d}|j,\Upsilon(\vec n)\rangle\no
&=\int \prod_{a=1}^5 \dd g_a\, \delta_{ij} Q_{na}^i Q_{nb}^j \delta_{kl} Q_{mc}^k Q_{md}^l {\prod_{cd}}' P^{cd}\;,
\label{eq:QQP}
\end{align}
\fi
\end{widetext}
where the product $\prod'$ is over couples $(cd)$ different from $(na),(nb),(mc),(md)$.

Now let us come to express the insertion $Q_{ab}^i$ in \eqref{Qab bracket} as a group integral.
Using the invariance properties of the map $Y$
\begin{equation}
Y J_{ab}^i|j_{ab},m_{ab}\rangle=J^i_{ab}Y|j_{ab},m_{ab}\rangle\;
\end{equation}
and the fact that the generator $J^i_{ab}$ of $SU(2)$  can be obtained as the derivative
\begin{equation}
i\frac{\partial}{\partial \alpha^i}\Big|_{\alpha^i=0}\big(e^{-i\alpha^i\tau_i}\big)=J^i_{ab\pm},
\end{equation}
we have that
\begin{widetext}
\begin{align}
Q_{ab}^i\equiv& \langle j_{ab},-\vec{n}_{ab}(\xi)|\,Y^\dag g_a^{-1} g_b Y (E^a_b)^i |j_{ab},\vec{n}_{ba}(\xi)\rangle\nonumber\\
=&i\gamma\frac{\partial}{\partial \alpha^i}\Big|_{\alpha_i=0}
\int_{\CP^1}\dd\bz\;\overline{g_a\,f^{j_{ab}}_{J\xi_{ab}}(z)^{(j_{ab},\gamma j_{ab})}}\;g_b\;e^{-i\alpha_i\tau_i}\, f^{j_{ab}}_{\xi_{ba}}(z)^{(j_{ab},\gamma j_{ab})}  \nonumber\\
=&i\gamma\frac{\partial}{\partial \alpha^i}\Big|_{\alpha_i=0}\int_{\CP^1}\frac{d_{j_{ab}}}{\pi}\frac{-\dd\bz_{ab}}{\la Z_{ab},Z_{ab}\ra \la \widetilde{Z}_{ba},\widetilde{Z}_{ba}\ra}%\;\no
%&\qquad\qquad\qquad
\left(\frac{\la \widetilde{Z}_{ba},\widetilde{Z}_{ba}\ra}{\la Z_{ab},Z_{ab}\ra}\right)^{i\gamma j_{ab}}\left(\frac{\la J\xi_{ab},Z_{ab}\ra^2\la \widetilde{Z}_{ba},\xi_{ba}\ra^2}{\la Z_{ab},Z_{ab}\ra \la \widetilde{Z}_{ba},\widetilde{Z}_{ba}\ra}\right)^{j_{ab}}\label{Qab_halfcalculate}
\end{align}
\end{widetext}
where
\be
\widetilde{Z}_{ba}=(g_b\;e^{-i\alpha_i\tau_i})^{\dag}z_{ab}=e^{-i\alpha_i\tau_i}\,Z_{ba}
\ee
Using
\be
\la\widetilde{Z}_{ba},\widetilde{Z}_{ba}\ra=\la Z_{ba},Z_{ba}\ra
\ee
equation (\ref{Qab_halfcalculate}) turns out to be
\begin{widetext}
\begin{align}
Q_{ab}^i=&i\gamma\frac{d_{j_{ab}}}{\pi}\int_{\CP^1}\dd\tilde{\bz}_{ab}\;\left(\frac{\la Z_{ba},Z_{ba}\ra}{\la Z_{ab},Z_{ab}\ra}\right)^{i\gamma j_{ab}}\frac{\partial}{\partial \alpha^i}\Big|_{\alpha_i=0}\left(\frac{\la J\xi_{ab},Z_{ab}\ra^2\la e^{-i\alpha_i\tau^i}{Z}_{ba},\xi_{ba}\ra^2}{\la Z_{ab},Z_{ab}\ra \la Z_{ba},Z_{ba}\ra}\right)^{j_{ab}}\nn\\
=&i\gamma\frac{d_{j_{ab}}}{\pi}\int_{\CP^1}\dd\tilde{\bz}_{ab}\;\left(\frac{\la Z_{ba},Z_{ba}\ra}{\la Z_{ab},Z_{ab}\ra}\right)^{i\gamma j_{ab}}\left(\frac{\la J\xi_{ab},Z_{ab}\ra^2\la {Z}_b,\xi_{ba}\ra^2}{\la Z_{ab},Z_{ab}\ra \la Z_{ba},Z_{ba}\ra}\right)^{j_{ab}}\no
&\qquad\qquad\frac{2j_{ab}}{\la Z_{ba},\xi_{ba}\ra}
\frac{\partial}{\partial \alpha^i}\Big|_{\alpha_i=0}\la e^{-i\alpha_i\tau^i}\,Z_{ba},\xi_{ba}\ra\nn\\
=&\frac{d_{j_{ab}}}{\pi}\int_{\CP^1}\dd\tilde{\bz}_{ab}\;\left(\frac{\la Z_{ba},Z_{ba}\ra}{\la Z_{ab},Z_{ab}\ra}\right)^{i\gamma j_{ab}}\left(\frac{\la J\xi_{ab},Z_{ab}\ra^2\la {Z}_b,\xi_{ba}\ra^2}{\la Z_{ab},Z_{ab}\ra \la Z_{ba},Z_{ba}\ra}\right)^{j_{ab}}
2\gamma j_{ab}\frac{\la \tau^i Z_{ba},\xi_{ba}\ra}{\la  Z_{ba},\xi_{ba}\ra}.\label{Qab:calculation}
\end{align}
\end{widetext}
Let
\be
A_{ab}^i\equiv\gamma j_{ab}\frac{\la \sigma^i Z_{ba},\xi_{ba}\ra}{\la  Z_{ba},\xi_{ba}\ra}\label{Aab}.
\ee
and use $K_{ab}$  given by equation (\ref{Kab}), one can rewrite equation \Ref{Qab:calculation} simply as
\be
Q_{ab}^i=\int\frac{d_{j_{ab}}}{\pi}\dd\tilde{\bz}_{ab}K_{ab}(A_{ab})^i.\label{Qab}
\ee
This equation can be used to rewrite equations (\ref{eq:QQP}) as
\begin{widetext}
\begin{align}
&\bra{W}E^a_n\cdot E^b_n\ket{j,\Upsilon(\bn)}=\int\prod_{a=1}^4\dd g_a\, \int \left(\prod_{a'<b'}\;\frac{d_{j_{a'b'}}}{\pi}\dd \tilde{\bz}_{a'b'}\right)\;q_n^{ab}\,e^{S}\\
&\bra{W}E^a_n\cdot E^b_n\ E^c_m\cdot E^d_m\ket{\Upsilon(\bn)}=\int\prod_{a=1}^4\dd g_a\, \int \left(\prod_{a'<b'}\;\frac{d_{j_{a'b'}}}{\pi}\dd \tilde{\bz}_{a'b'}\right)\;q_n^{ab}q_n^{cd}\,e^{S}
\end{align}
\end{widetext}
with
\be
q_n^{ab}\equiv A^a_n\cdot A^b_n.
\ee
Here we remove a redundant $\d g_5$ integral, as discussed in Sec. \ref{eprl}.
Then the two-point function (\ref{super:Gnm}) can be reexpressed in terms of group integrals:
\begin{widetext}
\begin{equation}
G_{nm}^{abcd}=\frac{\sum_j\psi_j\int \d^4g\,\d^{10} z\; q_n^{ab}q_m^{cd}e^{S}}{\sum_j\psi_j\int \d^4g\,\d^{10} z\; e^{S}}-
\frac{\sum_j\psi_j\int \d^4g\,\d^{10} z\; q_n^{ab}e^{S}}{\sum_j\psi_j\int \d^4g\,\d^{10} z\; e^{S}}\;
\frac{\sum_j\psi_j\int \d^4g\,\d^{10} z\; q_m^{cd}e^{S}}{\sum_j\psi_j\int \d^4g\,\d^{10} z\; e^{S}},\label{integ:Gnm}
\end{equation}
\end{widetext}
where the group integral is over $\d^4g=\prod_{a=1}^4\d g_a$ and $\d^{10} z$ is short for the integral measure $\prod_{a<b}\;\frac{d_{j_{ab}}}{\pi}\dd \tilde{\bz}_{ab}$ over $\cp$.

%-------------------------------------------------------------------
%  Lorentzian geometry
%-------------------------------------------------------------------
%---------------------------------------------------------------------------------------
% Asymptotic
%---------------------------------------------------------------------------------------
\section{Asymptotic expansion of the two-point function for large spin}\label{sec:propagator_asymptotic}
In this section we study the large-$j_o$ asympototics of the correlation function (\ref{integ:Gnm}).  We use the technique developed in \cite{Bianchi:2009ri}. The idea is rescaling the spins  $j_{ab}$ and $(j_o)_{ab}$ by an integer $\lambda$ so that $j_{ab}\rightarrow\lambda j_{ab}$ and $(j_o)_{ab}\rightarrow\lambda (j_o)_{ab}$, thus the two-point function (\ref{integ:Gnm}) can be reexpressed as $G(\lambda)$. To study large-spin limit turns then to study large-$\lambda$ limit, via stationary phase approximation. In Sec. \ref{sec:spa} we give a brief framework of this technique. Then in Sec. \ref{sec:crit}-\ref{sec:hessian} we give the detailed calculation.

\subsection{The rescaled correlation function and stationary phase approximation}\label{sec:spa}
As in \cite{Bianchi:2009ri}, let $j_{ab}\rightarrow\lambda j_{ab}$ and $(j_o)_{ab}\rightarrow\lambda (j_o)_{ab}$, we rescale the correlation function (\ref{integ:Gnm}) as
\begin{widetext}
\begin{equation}
G_{nm}^{abcd}(\lambda)=\frac{\sum_j\int\ d^4g\,\d^{10} z\, q_n^{ab}q_m^{cd}e^{\lambda {\tot}}}{\sum_j\int\ d^4g\,\d^{10} z\,  e^{\lambda {\tot}}}-
\frac{\sum_j\int\ d^4g\,\d^{10} z\,  q_n^{ab}e^{\lambda {\tot}}}{\sum_j\int\ d^4g\,\d^{10} z\,  e^{\lambda {\tot}}}\;
\frac{\sum_j\int\ d^4g\,\d^{10} z\,  q_m^{cd}e^{\lambda {\tot}}}{\sum_j\int\ d^4g\,\d^{10} z\,  e^{\lambda {\tot}}},\label{rescale:Gnm}
\end{equation}
\end{widetext}
 where the ``total action'' is defined as $S_{\text{tot}}=\log \psi+ S$ or more explicitly as
 \begin{widetext}
\begin{align}
S_{\mathrm{tot}}(j,g,\bz)=\;-\frac{1}{2}\sum_{ab,cd} \gamma\alpha^{(ab)(cd)}\, \frac{j_{ab}-(j_o)_{ab}}{\sqrt{(j_o)_{ ab}}}\,\frac{j_{cd}-(j_o)_{ cd}}{\sqrt{(j_o)_{ cd}}}-i\sum_{ab}\gamma \phi_o^{ab}\,(j_{ab}-(j_o)_{ ab})+S(j,g,\bz)\;\label{stot}
\end{align}
\end{widetext}
Using Euler-Maclaurin formula, one can evaluate the sums over spins $j$ using integrals in the large $\lambda$ limit:

\begin{align}
\sum_j   \;q_n^{ab} \, e^{\lambda S_{\text{tot}}}=&\int \dd^{10}j\, \,q_n^{ab}\, e^{\lambda S_{\text{tot}}}+O(\lambda^{-N})\no
&\qquad \forall N>0\;,
\end{align}

so that the rescaled correlation function (\ref{rescale:Gnm}) can be approximately expressed in the large $\lambda$ limit
\begin{widetext}
\begin{equation}
G_{nm}^{abcd}(\lambda)=\frac{ \int\d^{10}j\,\ d^4g\,\d^{10} z\, q_n^{ab}q_m^{cd}e^{\lambda {\tot}}}{ \int\d^{10}j\,\ d^4g\,\d^{10} z\,  e^{\lambda {\tot}}}-
\frac{ \int\d^{10}j\,\ d^4g\,\d^{10} z\,  q_n^{ab}e^{\lambda {\tot}}}{ \int\d^{10}j\,\ d^4g\,\d^{10} z\,  e^{\lambda {\tot}}}\;
\frac{ \int\d^{10}j\,\ d^4g\,\d^{10} z\,  q_m^{cd}e^{\lambda {\tot}}}{ \int\d^{10}j\,\ d^4g\,\d^{10} z\,  e^{\lambda {\tot}}}.\label{integ:j}
\end{equation}
\end{widetext}
We will study the large-$\lambda$ asymptotics of expression (\ref{integ:j}).

Let us rewrite the two-point function (\ref{integ:j}) formally as
\begin{widetext}
\begin{equation}
G(\lambda)=\frac{\int\dd x\,  \,p(x)q(x)\,e^{\lambda S(x)}}{\int\dd x\,  \,e^{\lambda S(x)}}-\frac{\int\dd x\,  \,p(x)\,e^{\lambda S(x)}}{\int\dd x\,  \,e^{\lambda S(x)}} \frac{\int\dd x\,  \,q(x)\,e^{\lambda S(x)}}{\int\dd x\,  \,e^{\lambda S(x)}}\;,
\end{equation}
\end{widetext}
then the asymptotic expansion of $G(\lambda)$ for large $\lambda$ is given by
\begin{equation}
G(\lambda)=\frac{1}{\lambda}\,(H^{-1})^{ij}\,p'_i(x_o)\,q'_j(x_o)\,+\mathcal O({\textstyle\frac{1}{\lambda^2}})\;.\label{eq:connected formula}
\end{equation}
Here $x_o$ is the \emph{critical point}, i.e. the stationary point where the real part of the action vanishes, $\mathrm{Re}S(x_o)=0$; $p'_i=\partial p/\partial x^i$, $H$ is the Hessian matrix at the critical point $H=S''(x_o)$. Our task is to obtain the critical point, the derivative of the insertions in Sec. \ref{sec:crit} and the Hessian in Sec. \ref{sec:hessian}.

\subsection{The critical point and the derivative of insertions}\label{sec:crit}
The critical point is the one where the real part and the derivatives of the total action $\tot$ vanish.
The real part of the total action (\ref{stot}) is given by
\begin{widetext}
\begin{align}
\text{Re} S_{\text{tot}}=\;-\sum_{ab,cd} \gamma(\text{Re}\, \alpha)^{(ab)(cd)}\, \frac{j_{ab}-(j_o)_{ab}}{\sqrt{(j_o)_{ab}}}\,\frac{j_{cd}-(j_o)_{ cd}}{\sqrt{(j_o)_{ cd}}}+\sum_{(ab)} j_{ab} \log \frac{ |\la J\xi_{ab},Z_{ab} \ra |^2 |\la Z_{ba},\xi_{ba} \ra |^2}{\la \z_{ab}, \z_{ab} \ra \la \z_{ba}, \z_{ba} \ra} \;.
\end{align}
\end{widetext}
Having assumed that the matrix $\alpha$ in the boundary state has positive definite real part, we have
 that the real part of the total action is negative or vanishing,  $\text{Re} S_{\text{tot}}\leq 0$. In particular the total action vanishes for the configuration of spins $j_{ab}$ and group elements $g$ satisfying
\begin{subequations}\label{max}
\be
j_{ab}=(j_o)_{ab}\;,\label{eq:j critical}
\ee
\begin{align}
J\xi_{ab} =& \frac{e^{i \phi_{ab}}}{\parallel \z_{ab} \parallel} Z_{ab},  \quad  \xi_{ba} = \frac{e^{i \phi_{ba}}}{\parallel
 \z_{ba} \parallel} Z_{ba},
\label{eq:g critical}
\end{align}
\end{subequations}
where
 $\parallel \z_{ab} \parallel$ is the norm of $\z_{ab}$ induced by the Hermitian inner product, and $\phi_{ab}$ and $\phi_{ba}$ are phases.

The requirement that the variations of the total action with respect to the spinors $z_{ab}$ and $\zb_{ab}$ vanishes, $\delta_z S_{\text{tot}}=\delta_{\zb}S_{\text{tot}}=0$, lead both to
\begin{align}
e^{-i\phi_{ab}}\frac{g_aJ\xi_{ab}}{\|Z_{ab}\|}=e^{-i\phi_{ba}}\frac{g_b\xi_{ba}}{\|Z_{ba}\|}\label{critical:z}
\end{align}
evaluated at the maximum point (\ref{eq:g critical}). For the group variables, $\delta_{g}\,\tot=0$ leads to
  \be
  \sum_{b : b \neq a} j_{ab} \mathbf{n}_{ab} = 0
  \ee
evaluated at the maximum point (\ref{eq:g critical}).
In fact the normals $\vec{n}_{ab}$ in the boundary state are chosen to satisfy the closure condition at each node. Therefore the critical points in the group variables are given by all the solutions of equation (\ref{eq:g critical}).

The variations of the total action with respect to the spins $j$ turns out to be
\begin{align}
    \frac{\partial S_{\text{tot}}}{\partial j_{ab}}=-\sum_{cd}\,\frac{\gamma\alpha^{(ab)(cd)}(j_{cd}-(j_o)_{ cd})}{\sqrt{(j_o)_{ ab}}\sqrt{(j_o)_{ cd}}}-i \gamma\phi_o^{ab}+\frac{\partial S}{\partial j_{ab}}\;.
\label{eq:j stationary}
  \end{align}
Imposing the maximal-point equation (\ref{max}) and the critical-point equation (\ref{critical:z}), equation (\ref{eq:j stationary}) is reduced into
\begin{align}
\frac{\partial S_{\text{tot}}}{\partial j_{ab}}\Big|_{\mathrm{crit}}=-i\gamma\phi_o^{ab}+i\mu S_{\mathrm{Regge}},\label{mu=1}
\end{align}
  where the parameter $\mu=\pm1$ measures  the discrepancy of the orientations of the 4-simplex $\sigma$: there are two orientations of the 4-simplex $\sigma$, one inherited from the Minkowski space where $\sigma$ is embedded, the other induced from the boundary data; $\mu=1$ if these two agree and $\mu=-1$ otherwise. The requirement that equation (\ref{mu=1}) vanishes selects $\mu=1$, which means we only consider the case when the orientation of the boundary data agrees with the one induced from the Minkowski space.

We end this subsection by the first derivative of the insertion $q_n^{ab}(g,z)$ evaluated at the critical point:
\begin{widetext}
\begin{align}\label{derivative insertion}
\delta_{z_{an}} q_{n}^{ab}\Big|_{\mathrm{crit}}
=&0\\
\delta_{\zb_{an}} q_{n}^{ab}\Big|_{\mathrm{crit}}
=&\gamma^2(j_o)_{na}(j_o)_{nb}\,\left(
\frac{g_a\sigma^i\xi_{an}}{\|Z_{an}\|}n^i_{bn}
-\frac{g_a\xi_{an}}{\zn{an}}\vec{n}_{an}\cdot \vec{n}_{bn}
\right)\\
\delta^r_{g_a} q_n^{ab}\Big|_{\mathrm{crit}}=&\gamma^2(j_o)_{na}(j_o)_{nb}\,
\left(
n^i_{bn}\ip{\xi_{an},\vec{L}\sigma_i\xi_{an}}-\frac{i}{2}(\vec{n}_{an}\cdot \vec{n}_{bn})\vec{n}_{an}
\right)\\
\delta^b_{g_a} q_n^{ab}\Big|_{\mathrm{crit}}=&\gamma^2(j_o)_{na}(j_o)_{nb}\,
\left(
n^i_{bn}\ip{\xi_{an}\vec{K}\sigma_i\xi_{an}}-\frac{1}{2}(\vec{n}_{an}\cdot \vec{n}_{bn})\vec{n}_{an}
\right)\\
\delta_{j_{cd}}q_{n}^{ab}\Big|_{\mathrm{crit}}=&\gamma^2\delta_{j_{cd}}\Big|_{(j_o)_{ab}}
(j_{an}\vec{n}_{an}\cdot j_{bn}\vec{n}_{bn})\label{derivative insertions}
\end{align}
\end{widetext}

\subsection{Hessian matrix of the total action}\label{sec:hessian}
Following the stationary phase approximation introduced in section \ref{sec:spa}, once the Hessian matrix is obtained, one can get asymptotic expansion of the two-point function (\ref{integ:Gnm}) by using equations (\ref{eq:connected formula}) and (\ref{derivative insertion}). Now let us come to calculate the Hessian matrix.

The Hessian is defined as the matrix of the second derivatives of the total action where the variable $g_5$ has been gauge fixed to the identity.
We split the Hessian matrix into derivatives w.r.t. the spins $j$,w.r.t. the group elements $g$ and w.r.t. $z$. The Hessian will then be a
$(10+24+20)\times (10+24+20)$ matrix
\be\label{hessian}
\tot^{''}=\left(
\begin{array}{cccc}
    Q_{jj}  &    0_{10\times24}        &    0_{10\times20}\\
    0_{24\times10}  &    H_{gg}      &    H_{gz}\\
    0_{20\times10}  &    H_{zg}  &    H_{zz}
\end{array}
\right)
\ee
as
\begin{align}
\delta_j\delta_gS_{\tot}=0\quad \delta_j\delta_zS_{\tot}=0.
\end{align}

We will now describe the non-vanishing blocks of this matrix. $Q_{jj}$ is a $10\times 10$ matrix containing only derivatives with respect to the spins $j_{ab}$, with elements
\begin{widetext}
\begin{align}
Q_{(ab)(cd)}=\delta_{j_{ab}}\delta_{j_{cd}}\tot\Big|_{\mathrm{crit}}=&
-\frac{\gamma\alpha^{(ab)(cd)}}{\sqrt{(j_o)_{ab}}\sqrt{(j_o)_{cd}}}+\delta_{j_{ab}}\delta_{j_{cd}}\Big|_{\mathrm{crit}}S_{\mathrm{Regge}}.
\end{align}
\end{widetext}

$H_{gg}$ is a $(4\times 6)\times(4\times6)$ matrix containing only
derivatives with respect to the group elements $g_a$. Note that due to the form of the action, derivatives with
respect to two different group variables will be zero and it will be block diagonal
\begin{align}
H_{gg}=\left(
\begin{array}{cccc}
    H_{11}    & 0           &  0          & 0  \\
    0              & H_{22} &  0          & 0  \\
    0              & 0           & H_{33} & 0  \\
    0              & 0           &  0          &    H_{44}\\
\end{array}
\right)
\end{align}
Each $H_{aa}$ is a $6\times6$ matrix. The variation has been performed by splitting the $\slc$
element into a boost and a rotation generator. This gives
\be
H_{aa} =
\left(
\begin{array}{cc}
    H^{rr}_{(ai)(aj)}    & H^{br}_{(ai)(aj)}         \\
    H^{rb}_{(ai)(aj)}    & H^{bb}_{(ai)(aj)}     \\
\end{array}
\right)
\ee
with $3\times3$ matrices
\begin{align}
\h^{rr}=&\half\sum_{b:b\neq a}j_{ab}(-\delta^{ij}+n^in^j+i\epsilon^{ij}{}_kn^k)\\
\h^{rb}=&-\frac{i}{2}\sum_{b:b\neq a}j_{ab}(-\delta^{ij}+n^in^j+i\epsilon^{ij}{}_kn^k)\\
\h^{br}=&-\frac{i}{2}\sum_{b:b\neq a}j_{ab}(-\delta^{ij}+n^in^j+i\epsilon^{ij}{}_kn^k)\\
\h^{bb}=&2\sum_{b:b\neq a}j_{ab}(1+\frac{i}{2}\gamma)\Big(-\delta^{ij}+n^in^j+i\epsilon^{ij}{}_kn^k\Big).
\end{align}

$H_{zz}$ is a matrix containing only derivatives with respect to the spinors $z_{ab}$ and $\bar{z}_{ab}$.
\begin{align}
H_{zz}=\left(
\begin{array}{cc}
    S^{''}_{zz}    & S^{''}_{z\zb}         \\
    S^{''}_{\zb z}    & S^{''}_{\zb\zb}     \\
\end{array}
\right)
\end{align}
Since each spinor $z_{ab}$ has two compononts $(z_o\quad z_1)$, each block $S^{''}$ of $H_{zz}$ seems to be $20\times 20$ matrix and thus $H_{zz}$ seems to be $40\times40$; however, this $40\times40$ matrix is degenerate due to the homogeneity of the representation functions (\ref{homofunc}) which we discuss in section \ref{eprl}. Remind that although we still keep the formulae of $f(\bf{z})$, we have chosen a section for $(z_o\quad z_1)$ as $(z_o/z_1\quad 1)$.  Hence the matrix $H_{zz}$ reduced to be $20\times20$ and non-degenerate, by removing the derivatives with respect to the second component $z_1$:
\begin{align}
H_{zz}=\left(
\begin{array}{cc}
    0_{10\times10}    & S^{''}_{z\zb}         \\
    S^{''}_{\zb z}    & 0_{10\times10}     \\
\end{array}
\right)
\end{align}
with $S^{''}_{zz}$ and $S^{''}_{\zb\zb}$ vanishing, $S^{''}_{z\zb}$ and $S^{''}_{z\zb}$ diagonal matrices. The diagonal elements are first components of the matrix
\begin{widetext}
\begin{align}
H_{z_{ab}\zb_{ab}}&={j_{ab}}\Big(\frac{2(g_aJ\xi_{ab})(g_aJ\xi_{ab})^{\dag}+(i\gamma-1)g_ag_a^{\dag}}{\|Z_{ab}\|}
-\frac{(i\gamma+1)g_bg_b^{\dag}}{\|Z_{ba}\|}
\Big)
\end{align}
\end{widetext}

$H_{gz}=\left(S^{''}_{gz}\qquad S^{''}_{g\zb} \right)$ is a $24\times20$ matrix, containing derivatives with respect to spinor $z_{ab}$ and group element $g_a$. Again, we consider the reduced matrix, by removing the derivatives with respect to the sencond component of spinor $z_{ab}$. The non-vanish elements are the first components of
\begin{widetext}
\begin{align}
    H^{rz}_{(ai)(ab)}&=e^{i\phi_{ab}}\frac{j_{ab}}{\|Z_{ab}\|}\Big(
    (1-i\gamma)(L^ig_aJ\xi_{ab})^{\dag}-i(1+\gamma)\nb_{ab}(g_aJ_{\xi_{ab}})^{\dag}
    \Big)\nonumber\\
    H^{bz}_{(ai)(ab)}&=e^{i\phi_{ab}}\frac{j_{ab}}{\|Z_{ab}\|}\Big(
    (1-i\gamma)(K^ig_aJ\xi_{ab})^{\dag}-i(1+\gamma)\nb_{ab}(g_aJ_{\xi_{ab}})^{\dag}
    \Big)\nonumber\\
   H^{r\zb}_{(ai)(ab)}&=-e^{-i\phi_{ab}}\frac{j_{ab}}{\|Z_{ab}\|}
    (1+i\gamma)(\nb_{ab}+L^ig_a)\jx\nonumber\\
    H^{b\zb}_{(ai)(ab)}&=-e^{-i\phi_{ab}}\frac{j_{ab}}{\|Z_{ab}\|}
    (1+i\gamma)(\nb_{ab}+K^ig_a)\jx\nonumber
\end{align}
\end{widetext}

\subsection{The $\gamma\rightarrow0$ limit}
Now we obtain the Hessian matrix. This matrix is very complicated and its inverse is not easy to express explicitly. Despite this difficulty, it is possible to look at the approximate inverse in a proper limit, where the Barbero-Immirzi parameter is taken to zero $\gamma\to 0$, and the spin of the boundary state is taken to infinity $j\to \infty$, keeping the size of the quantum geometry $A\sim \gamma j$ finite and fixed. In this limit, the Euclidean two-point \cite{Bianchi:2009ri} as well as three-point \cite{rovelli:2011kf}  correlation functions matches the ones in Regge calculus \cite{Bianchi:2007vf}. More recently, Magliaro and Perini propose in \cite{Magliaro:2011dz,Magliaro:2011zz} to understand this limit as the continuous spectrum limit of the fundamental geometric operators, since the Barbero-Immirzi parameter $\gamma$ enters the discrete spectra of area and volume
operators \cite{Rovelli:1994ge,Ashtekar:1996eg,Ashtekar:1997fb,Engle:2007mu,Ding:2009jq,Ding:2010ye}. Now let us come to study the Hessian matrix in this limit and the correlation function as well.

Before we take this limit, let us comment more on $j\rightarrow\infty$. In the Lorentzian signature, we do not have the regular 4-simplex, so that we do not have $j_{ab}=j$ or $(j_o)_{ab}=j_o$ for all triangles. Since all $j_{ab}\rightarrow\infty$ in the large-$j$ limit, however, we can factorize each spin into $j_{ab}=j \epsilon_{ab}$ and let $j\rightarrow\infty$ and $\epsilon_{ab}$ finite. Thus we have a fundamental area $A=\gamma j$ and $\epsilon_{ab}$ is obtained when we use the fundamental area $A=\gamma j$ to measure triangle $(ab)$.

Now let us take the limit for the Hessian matrix.
\begin{itemize}
  \item The derivatives with respect to group elements $H_{gg}=jH^{\epsilon}_{gg}$, with $24\times24$ matrix $H^{\epsilon}_{gg}$ independent from $j$ and $\gamma$.
  \item The derivatives with respect to spinors $H_{zz}=j(H^{\epsilon}_{zz}+\mathcal{O}(\gamma))$, with $20\times20$ matrix $H^{\epsilon}_{zz}$ independent from $j$ and $\gamma$.
    \item The derivatives with respect to group elements and spinors $H_{gz}=j(H^{\epsilon}_{gz}+\mathcal{O}(\gamma))$, with $24\times20$ matrix $H^{\epsilon}_{gz}$ independent from $j$ and $\gamma$.
\end{itemize}
Thus we can express the Hessian matrix in the limit as
\be\label{hessian limit}
\tot^{''}=\left(
\begin{array}{ccc}
    Q_{jj}  &    0_{10\times44}        \\
    0_{44\times10}  &    j(H^{\epsilon}+\mathcal{O}(\gamma))
\end{array}
\right)
\ee
with $44\times44$ matrix $H^{\epsilon}$ independent from $j$ and $\gamma$. The inverse of this Hessian matrix (\ref{hessian limit}) is simply
\begin{align}
(\tot^{''})^{-1}=\left(
\begin{array}{ccc}
    Q^{-1}_{jj}  &    0_{10\times44}        \\
    0_{44\times10}  &    j^{-1}((H^{\epsilon})^{-1}+\mathcal{O}(\gamma))
\end{array}
\right)
\end{align}

Substituting (\ref{derivative insertion})-(\ref{derivative insertions}) and the approximate hessian matrix (\ref{hessian limit})  into (\ref{eq:connected formula}) we obtain the asymptotic expansion of the correlation function
\begin{align}
G_{nm}^{abcd}=&\sum_{p,q,r,s} Q^{-1}_{(pq)(rs)}\;\frac{\partial q_n^{ab}}{\partial j_{pq} }\; \frac{\partial q_m^{cd}}{\partial j_{rs}}\;+\no
&\;+\gamma^4j^3(X^{\epsilon}+\mathcal{O}(\gamma))
\end{align}
with $X^{\epsilon}$ independent from the boundary spin $j$ and from $\gamma$, as well as the combination
\begin{align}
R_{nm}^{abcd}=&\;\frac{1}{\gamma^3 j_o^3}\sum_{p<q,r<s} Q^{-1}_{(pq)(rs)}\;\frac{\partial q_n^{ab}}{\partial j_{pq} }\; \frac{\partial q_m^{cd}}{\partial j_{rs}}\;.
\end{align}
In terms of these quantities we have the following two-point correlation function
\begin{align}
G^{abcd}_{nm}(\alpha)=(\gamma j_o)^3(R_{nm}^{abcd}(\alpha)+O(\gamma))\label{correlation}
\end{align}

If we take the classical limit,
introduced in \cite{Bianchi:2009ri}, where the Barbero-Immirzi parameter is taken to zero $\gamma\to 0$, and the spin of the boundary state is taken to infinity $j\to \infty$, keeping the size of the quantum geometry $A\sim \gamma j$ finite and fixed \footnote{The finite and fixed area $A$ corresponds to the finite and fixed distance between the two points where the correlation functions are defined.}, the two-point function (\ref{correlation}) we obtain exactly matches the one obtained from Lorentzian Regge calculus \cite{Regge:1961px}.

%The correlation function is also calculated numerically in \cite{Neiman:2011gf}, and shown not to be invariant under under parity-odd permutations of equivalent nodes. Also, there is another generating functional technique, which is developed in \cite{Mikovic:2011zv,Mikovic:2011zx} can be applied to calculate correlation functions, which will be especially useful for computing $n$-point functions for larger $n$ and corrections.

Similarly, higher order $n$-point correlation functions can be computed using the generating functional technique discussed in \cite{Mikovic:2011zv,Mikovic:2011zx}.

\section{Conclusion}
In this paper we have studied in the Lorentzian signature the two-point correlation function of metric operators of EPRL spinfoam model.

The analysis presented involves two distinct ingredients.
The first is a setting for defining correlation functions. The setting is the boundary amplitude formalism. It involves a boundary semiclassical state $|\Psi_o\rangle$ which identifies the regime of interest, loop quantum gravity operators $E_n^a \cdot E_n^b$ which probe the quantum geometry on the boundary, a spin foam model $\langle W|$ which implements the dynamics. The formalism allows to define semiclassical correlation functions in a background-independent context.
The second ingredient consists in an approximation scheme applied to the quantity defined above. It involves a vertex expansion and a large spin expansion. It allows to estimate the correlation functions explicitly. The explicit result can then be compared to the correlation function of Regge calculus. In this paper we focused on the lowest order in the vertex expansion and the leading order in the large spin expansion.
In the following we collect some remarks on these two ingredients.

The boundary semiclassical state $\ket{\Psi_o}$ used here is \emph{Lorentzian}, in the sense that it is related to Lorentzian geometry of 4-simplex.  It is obtained via a superposition over spins of \emph{Lorentzian coherent spin network} $\ket{\Gamma_5,j_{ab},\Upsilon_a}$ with nodes labeled by \emph{Lorentzian Livine-Speziale coherent intertwiners}. The Lorentzian coherent intertwiner is the product of Livine-Speziale coherent intertwiner with a Lorentzian-geometry phase.

For the approximation scheme, we consider the limit, where the Barbero-Immirzi parameter is taken to zero $\gamma\to 0$, and the spin of the boundary state is taken to infinity $j\to \infty$, keeping the size of the quantum geometry $A\sim \gamma j$ finite and fixed. This limit corresponds to neglecting Planck scale discreteness and twisting effects, at large finite distances. It is interesting to notice that the same limit was considered in \cite{Bojowald:2001ep} in the context of loop quantum cosmology. In this limit, the two-point function we obtain exactly matches the one obtained from Lorentzian Regge calculus \cite{Regge:1961px}.

Deriving the LQG correlation function at the level of a single spin foam vertex is certainly only a first step. Within the setting of a vertex expansion, an analysis of the LQG correlation function for an arbitrary number of spinfoam vertices is needed. In \cite{Magliaro:2011dz,Magliaro:2011zz}, asymptotic analysis of spinfoams with an arbitrary number of vertices is studied in $\gamma\rightarrow0$ limit. Without taking $\gamma\rightarrow0$,  the large-$j$ of spinfoams with an arbitrary number of vertices is studied in \cite{Conrady:2008mk} with a closed manifold, and is generalized recently to the one with boundary \cite{Han:2011rf,Han:2011re}. It would be interesting to investigate the contribution of the $\gamma$-term to correlation functions when more than a single spin foam vertex is considered.

\section*{Acknowledgments}
We are grateful to Muxin Han, Frank Hellmann, Roberto Pereira, Elena Magliaro, Claudio Perini, Carlo Rovelli and Mingyi Zhang for useful comments. Y.D. is supported by CSC scholarship No. 2008604080.

\def\cprime{$'$}

\end{document}